# SoK: Systematic Analysis of Adversarial Threats Against Deep Learning Approaches for Autonomous Anomaly Detection Systems in SDN-IoT Networks

Tharindu Lakshan Yasarathna[a,], Nhien-An Le-Khac[a,]

[a]*School of Computer Science, University College Dublin, Belfield, Dublin, D04 V1W8, Ireland*

**Abstract**

Integrating Software Defined Networking (SDN) and the Internet of Things (IoT) enhances network control and flexibility. Deep Learning (DL)-based Autonomous Anomaly Detection (AAD) systems improve security by enabling real-time threat detection in SDN-IoT networks. However, these systems remain vulnerable to adversarial attacks that manipulate input data or exploit model weaknesses, significantly degrading detection accuracy. Existing research lacks a systematic analysis of adversarial vulnerabilities specific to DL-based AAD systems in SDN-IoT environments. This Systematisation of Knowledge (SoK) study introduces a structured adversarial threat model and a comprehensive taxonomy of attacks, categorising them into data-level, model-level, and hybrid threats. Unlike previous studies, we systematically evaluate white-box, black-box, and grey-box attack strategies across popular benchmark datasets (CICIDS2017, InSDN, and CICIoT2023). Our findings reveal that adversarial attacks can reduce detection accuracy by up to 48.4%, with Membership Inference causing the most significant drop. Carlini & Wagner and DeepFool achieve high evasion success rates. However, adversarial training enhances robustness, and its high computational overhead limits the real-time deployment of SDN-IoT applications. We propose adaptive countermeasures, including real-time adversarial mitigation, enhanced retraining mechanisms, and explainable AI-driven security frameworks. By integrating structured threat models, this study offers a more comprehensive approach to attack categorisation, impact assessment, and defence evaluation than previous research. Our work highlights critical vulnerabilities in existing DL-based AAD models and provides practical recommendations for improving resilience, interpretability, and computational efficiency. This study serves as a foundational reference for researchers and practitioners seeking to enhance DL-based AAD security in SDN-IoT networks, offering a systematic adversarial threat model and conceptual defence evaluation based on prior empirical studies.

*Keywords:* Software-Defined Networking, Internet of Things, Network Security, Deep Learning, Autonomous Anomaly Detection, Adversarial Attacks

## 1. Introduction

The fusion of Software Defined Networking (SDN) with the Internet of Things (IoT) has opened modern network architectures, enabling flexibility, scalability, and programmability. SDN centralises network control, allowing dynamic traffic management, while IoT connects billions of smart devices across various domains such as smart cities, healthcare, and industrial automation [27]. Despite these advancements, SDN-IoT networks are highly susceptible to cyber threats, including Distributed Denial-of-Service (DDoS) attacks, botnets, data breaches, and adversarial intrusions [27]. The increasing attack surface necessitates the deployment of Autonomous Anomaly Detection (AAD) systems, which leverage Deep Learning (DL) techniques to identify and mitigate malicious activities in real-time [17]. Unlike traditional rule-based or signature-based security mechanisms, AAD systems can adapt to new and evolving cyber threats without requiring constant human intervention [24, 17].

Although DL-based AAD systems enhance network security, they remain vulnerable to adversarial attacks, introducing subtle perturbations into input data to deceive the model. These attacks exploit inherent weaknesses in DL architectures, significantly degrading detection accuracy and allowing malicious traffic to bypass security mechanisms [44]. Adversarial threats in SDN-IoT networks are particularly concerning due to their dynamic and distributed nature, where real-time Anomaly Detection (AD) is crucial for maintaining network integrity. Attackers can manipulate DL-based AAD models through data poisoning, evasion attacks, and membership inference, leading to severe consequences, including misclassification of threats, data leaks, and large-scale security breaches [43]. Despite the critical impact of adversarial attacks on DL-based security systems, limited research has systematically analysed these vulnerabilities in SDN-IoT environments.

Existing studies on adversarial security in DL focus mainly on network intrusion detection systems (NIDS) and generic AI security models. Thus, the specific challenges and constraints of deploying AAD systems in SDN-IoT networks remain an open area of research. While prior research has explored some individual attack vectors or defensive mechanisms, a comprehensive systematisation of adversarial threats against DL-based AAD models is lacking. Furthermore, most studies fail to provide a structured threat model categorising adversarial attacks into data-level, model-level, and hybrid threats [33]. Additionally, the practical impact of adversarial attacks on DL-based AAD remains underexplored, as many evaluations rely on synthetic datasets or simplified models that do not accurately reflect the complexity of SDN-IoT networks. The lack of comparative analyses between white-box, black-box, and grey-box adversarial attack strategies further limits the applicability of existing research [44]. Moreover, defensive strategies such as adversarial training, feature squeezing, and model aggregation have been proposed. Still, they often introduce high computational overhead and fail to adapt dynamically to evolving attack strategies.

To address these critical gaps, this Systematisation of Knowledge (SoK) study presents a structured and comprehensive analysis of adversarial threats targeting DL-based AAD systems in SDN-IoT networks. Unlike previous works, we introduce a systematic adversarial threat model that categorises attacks into three distinct levels: data-level, model-level, and hybrid attacks. We empirically evaluate multiple adversarial attack techniques to assess their impact on DL-based AAD models. Our experimental findings reveal that adversarial attacks can significantly reduce detection accuracy and expose critical vulnerabilities in existing models. Beyond analysing adversarial threats, we evaluate existing defence mechanisms and identify their limitations in SDN-IoT environments. While countermeasures such as adversarial retraining and model ensembling provide partial protection, they are often insufficient against adaptive attack strategies. This study systematically categorises adversarial threats, conducts an in-depth impact assessment, and suggests practical recommendations for improving the resilience of DL-based AAD systems.

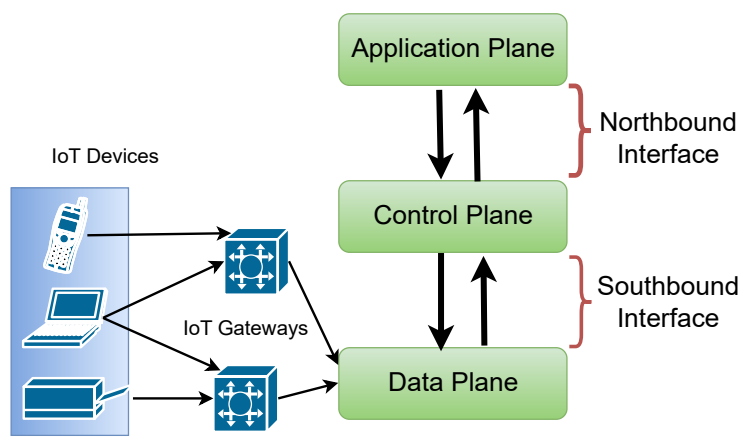

Figure 1: IoT-SDN Architecture that consists of application plane, control plane, and data plane

1.1. Methodology

This study follows a systematic experimental approach to assess the vulnerabilities of DL-based AAD models under adversarial settings and presents valuable suggestions for defences. We categorise adversarial threats into three primary types: data-level attacks, which involve direct manipulation of input data (e.g., Fast Gradient Method, poisoning attacks); model-level attacks, which exploit DL architecture weaknesses (e.g., Carlini & Wagner, DeepFool); and hybrid attacks, which combine data and model manipulations (e.g., transferable attacks, membership inference attacks). To evaluate these attacks, we conduct experiments using a Convolutional Neural Network (CNN) model trained on popular benchmark datasets, including CICIDS2017, InSDN, and CICIoT2023. We aim to systematically examine how adversarial attacks impact detection accuracy, computational performance, and model robustness in SDN-IoT environments. While CNNs are the focus of this study due to their efficiency in extracting spatial traffic-flow features, other DL architectures have also been employed for AAD in SDN-IoT networks. For instance, Recurrent Neural Networks (RNNs) and Long Short-Term Memory (LSTM) models capture temporal dependencies in sequential traffic data, while variants such as BiLSTM and AE-LSTM leverage bidirectional learning and autoencoding to enhance detection performance [12, 66]. We do not engage in direct comparative analysis across these models but instead use CNNs to demonstrate the extent of adversarial vulnerability in DL-based AAD systems. The key evaluation criteria include detection accuracy reduction—how adversarial attacks degrade classification performance—and computational cost analysis, which measures the overhead introduced by different attack strategies that bypass AAD mechanisms.

Additionally, we examine the effectiveness of various defensive mechanisms, such as adversarial training, feature squeezing, and model ensembling, to determine their ability to mitigate adversarial threats. The evaluation focuses on understanding how well these defences perform against attack types and the trade-offs between security enhancement and computational efficiency.

*1.2. Contributions*

The main contributions of this work can be summarised as follows:

- Analyse the characteristics and impact of adversarial attacks on DL-based AAD models, focusing on data, model, and hybrid attacks and their implications within the context of threat modelling in SDN-IoT environments.

- Conduct empirical experiments on popular benchmark datasets to assess detection accuracy, computational overhead, and model robustness.

- Based on empirical findings, provide practical recommendations to enhance the security of DL-based AAD systems in SDN-IoT networks.

Compared with prior surveys and SoK studies on adversarial machine learning, our work extends the state of the art in three ways. First, we evaluate a broader range of adversarial threat levels, including white-box, black-box, and grey-box scenarios, whereas earlier studies often focused on a single adversary type. Second, we propose a distinct taxonomy that systematically maps adversarial threats to data-level, model-level, and hybrid categories aligned with SDN-IoT architectural layers, providing a more transparent structure than existing works. Third, our study targets explicitly SDN-IoT environments, which remain underexplored in earlier adversarial ML research that primarily focused on general-purpose networks or image domains. In this way, our contributions complement and go beyond related studies by addressing both the unique characteristics and the open challenges of adversarial robustness in SDN-IoT systems.

*1.3. Paper Organisation*

The remainder of this paper is structured as follows: Section 2 provides background information on SDN-IoT security challenges and DL-based AAD models. Section 3 reviews related work, highlighting key gaps in adversarial security research. Section 4 introduces the systematic adversarial threat model and empirical findings on attack effectiveness. Section 5 discusses defensive strategies, evaluates existing mitigation techniques, and proposes novel countermeasures. Finally, Section 6 outlines limitations, future research directions, and Section 7 concludes the study by emphasising the need for more resilient and adaptive security mechanisms in DL-based AAD for SDN-IoT networks.

## 2. Background

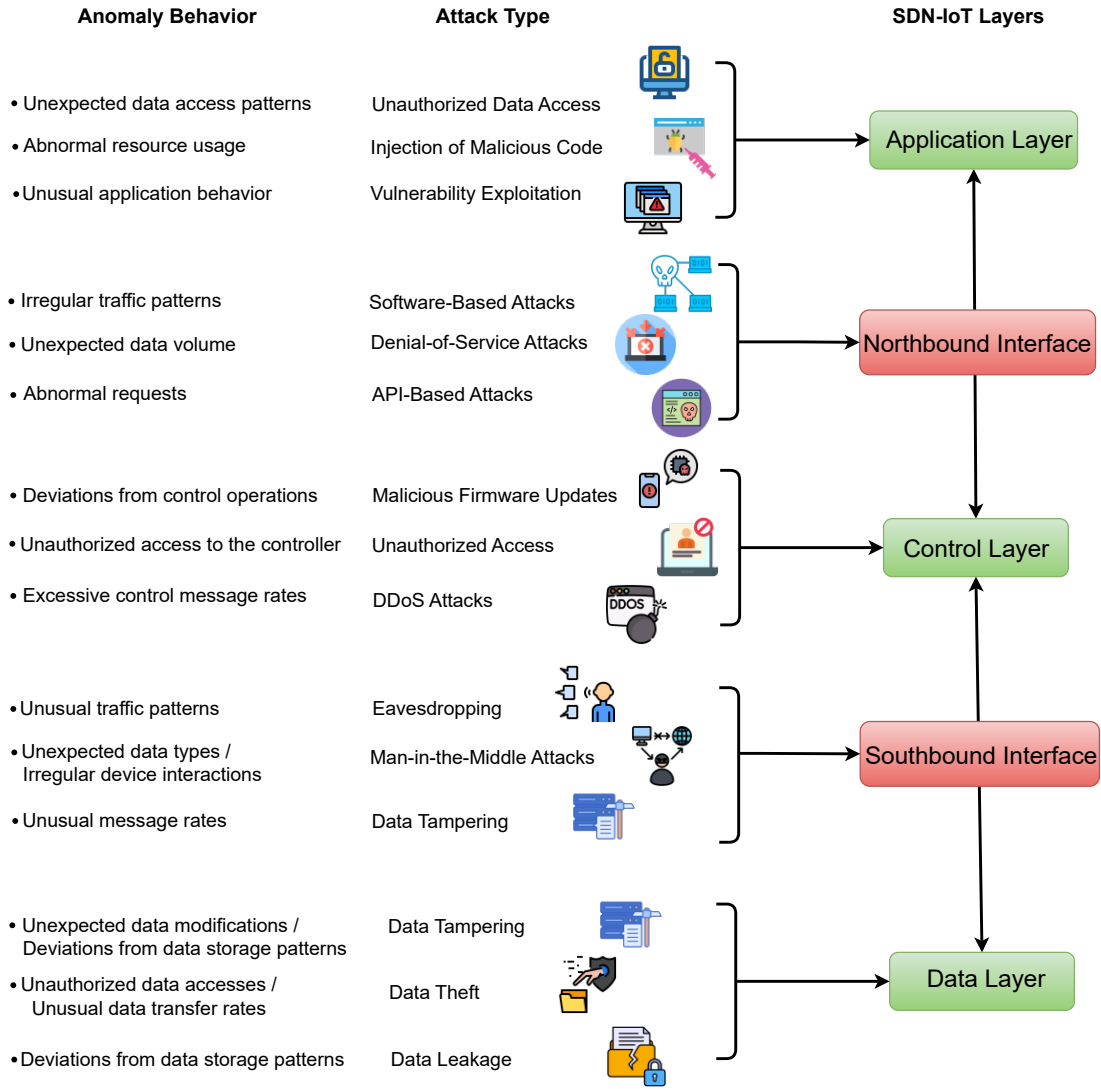

Figure 2: Mapping Anomaly Behaviours to Attack Types in IoT-SDN Architecture

*2.1. Evolution of SDN in IoT*

The Evolution of SDN in IoT has been transformative, reshaping the traditional network architecture. SDN introduces a centralised control plane that enables administrators to manage and optimise network resources dynamically [27]. In the context of IoT, this evolution addresses the challenges posed by the heterogeneity of devices, protocols, and communication patterns. SDN facilitates seamless communication and efficient resource allocation, laying the foundation for a highly responsive and adaptable network infrastructure [28].

In SDN-IoT, the network infrastructure is divided into three key components: the data plane, control plane, and application plane (see Figure 1). The data plane includes the IoT devices responsible for generating and transmitting data. The control plane, centrally managed by SDN controllers, orchestrates the traffic flow, making real-time routing and resource allocation decisions. Lastly, the application layer contains various IoT applications and services that leverage the orchestrated network to achieve specific objectives. This architecture enhances flexibility, allowing administrators to adjust network behaviour programmatically in response to changing requirements [37]. However, integrating diverse IoT devices and protocols introduces novel security challenges, necessitating innovative solutions to ensure the integrity and confidentiality of data transmitted within the SDN-IoT ecosystem [28].

## 2.2. Security Challenges in SDN-IoT

The proliferation of IoT devices within SDN introduces unique security challenges (see Figure 2). The diverse nature of connected devices and communication protocols creates vulnerabilities that traditional security measures could struggle to address. Threats such as unauthorised access, device vulnerabilities, data breaches, and network disruptions pose considerable risks [16]. Securing the SDN-IoT ecosystem requires a comprehensive approach to mitigate potential exploits and ensure the confidentiality, integrity, and availability of sensitive data. To address these challenges, integrating advanced technologies such as Artificial Intelligence (AI) and AD becomes crucial in enhancing the network's resilience and responsiveness to emerging threats [28, 37].

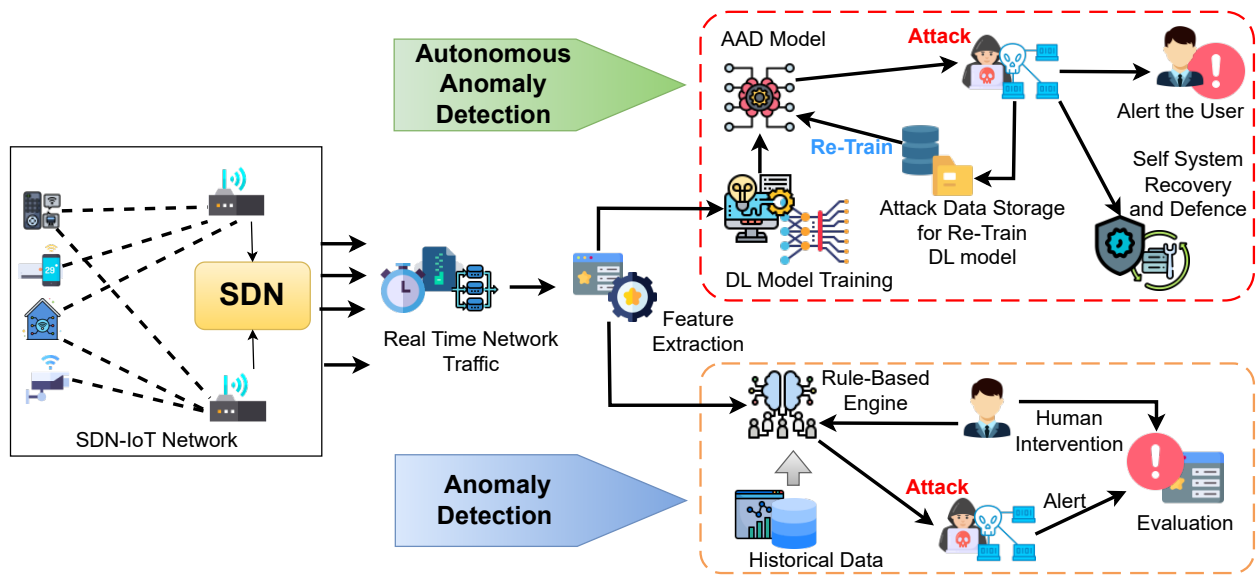

Figure 3: Comparison between Anomaly Detection and Autonomous Anomaly Detection in SDN-IoT networks

## 2.3. Autonomous Anomaly Detection

AAD involves detecting, investigating, and responding to anomalies without human intervention. It is essential in securing SDN-IoT networks, especially compared to traditional AD methods. AAD utilises DL algorithms for autonomous network behaviour analysis, distinguishing between normal and abnormal activities without relying on predefined signatures or extensive human intervention [17]. This capability is advantageous in addressing prevalent attack types in SDN-IoT, such as Distributed Denial of Service (DDOS) and Malware botnets. AAD further outshines traditional AD by demonstrating adaptability to evolving attack scenarios, particularly concerning data, model, and hybrid (data and model-level) attacks. Its adaptability empowers AAD to effectively discern abnormal patterns, establishing a robust defence mechanism against a spectrum of evolving attack contexts.

Notably, traditional AD may be more susceptible to adversarial manipulations with its rule-based approach [45]. Leveraging DL algorithms, AAD offers a proactive defence against adversarial attacks. Its autonomous and adaptable nature enhances resilience to adversarial manipulations, making it a more reliable choice in the face of evolving threats [24]. Considering AAD and AD, AAD's autonomous and adaptable characteristics make it well-suited for deployment, ensuring real-time threat detection and response. It makes AAD a crucial component in enhancing the security and responsiveness of SDN-IoT networks.

*2.4. Integration of DL-based AAD Models*

Integrating DL-based AAD models represents a strategic response to evolving cyber threats (see Figure 3). DL techniques are outstanding at learning intricate patterns from extensive datasets. Integrating DL techniques enhances anomaly detection capabilities in SDN-IoT networks by identifying slight and complex deviations. DL techniques can capture slight behaviours, effectively determining legitimate activities from potential security incidents [24]. However, this integration also introduces challenges related to the robustness of DL models against adversarial attacks, a critical consideration in ensuring the reliability of anomaly detection systems [27, 33, 30].

*2.5. Conventional and Adversarial Attacks: Key Success Factors*

Cyber threats in SDN-IoT networks can be categorised into conventional and adversarial attacks. Conventional attacks, such as Distributed Denial-of-Service (DDoS) attacks, malware infections, and unauthorised access, exploit vulnerabilities in network protocols, software, or human behaviour. These attacks typically rely on brute-force methods, credential theft, or known security loopholes and are often detected using rule-based or signature-based security systems [45]. In contrast, adversarial attacks manipulate input data to deceive DL models without altering the underlying network infrastructure. These attacks exploit model vulnerabilities rather than system flaws, introducing subtle perturbations that remain undetectable by traditional security measures but cause severe misclassification [33, 30]. Unlike conventional threats, adversarial attacks do not require brute force; instead, they exploit mathematical weaknesses in DL models, making them particularly dangerous in DL-based AAD in SDN-IoT environments.

Several key factors drive the success of adversarial attacks. High model sensitivity to small perturbations enables attackers to introduce minimal changes, which can significantly impact detection accuracy. Fragile decision boundaries in DL models make them vulnerable to misclassification, especially when adversarial examples are strategically placed near class boundaries. Lack of adversarial robustness in training means that models often fail to recognise adversarially crafted inputs, as they are primarily trained on clean datasets. Moreover, the transferability of adversarial examples enables attackers to craft inputs on one model and successfully deceive another, even in black-box scenarios where the model's structure and parameters are unknown. Information leakage through membership inference attacks further enhances adversarial success by allowing attackers to determine whether a specific data point was used during training, thereby aiding in the generation of targeted adversarial input [44]. Given the increasing reliance on DL for security in SDN-IoT networks, adversarial attacks pose a significant challenge [33]. Understanding their underlying mechanics and impact on AAD systems is crucial. This research highlights the urgent need for systematic threat modelling to assess the practical implications of adversarial attacks in dynamic and large-scale network environments.

## 3. Related work

Research on adversarial attacks against DL-based AAD systems in SDN-IoT networks has seen a surge in interest in recent years. Adversarial attacks exploit the vulnerabilities of DL models by introducing carefully crafted perturbations that can significantly degrade model performance, leading to severe security implications. Existing studies have primarily focused on theoretical aspects, empirical evaluations, and defence strategies, yet critical gaps remain in developing adaptive and scalable countermeasures.

Bai et al. [9] identified research gaps in adversarial robustness for continual learning settings and emphasised the importance of Task-Aware Boundary Augmentation (TABA) to mitigate adversarial attacks. Their study demonstrates that adversarial training can enhance model robustness; however, continual learning frameworks remain highly susceptible to adversarial forgetting, where previously learned robustness deteriorates over time when new tasks are introduced. It leads to critical challenges in dynamic SDN-IoT environments, where models must continuously adapt to evolving attack patterns without compromising their resilience to adversarial manipulations. Chakraborty et al. [15] conducted an extensive survey on adversarial attacks and defences in DL models, categorising adversarial strategies into white-box, black-box, and grey-box attacks. They highlight that DL models are particularly vulnerable in security-sensitive applications due to over-reliance on feature representations that attackers can subtly modify. Despite growing research on adversarial defences, existing methods, such as adversarial training, gradient masking, and ensemble learning, fail to provide holistic protection against adaptive adversarial strategies. Furthermore, their survey underscores the lack of adversarial benchmarks specific to SDN-IoT networks, which limits the evaluation of defence mechanisms in practical scenarios.

In contrast, a study [1] provided an overview of ML and DL algorithms for securing SDN environments. While their study focuses on traditional intrusion detection and prevention methods, it does not address the specific vulnerabilities of DL-based security mechanisms to adversarial attacks. Similarly, Alsoufi et al. [5] reviewed DL-based AD techniques in IoT security but did not consider how adversarial perturbations can compromise detection accuracy. In another work, Taheri et al. [49] explored DL applications for SDN security, emphasising the scalability and performance of DL-based approaches. However, these studies mostly overlook the threat posed by adversarial examples, particularly in real-time SDN-IoT scenarios where attackers can dynamically generate adversarial samples to evade detection. Aversano et al. [8] conducted a systematic review on DL approaches for IoT and SDN security, highlighting that while DL has achieved remarkable success in anomaly detection, most existing solutions remain vulnerable to adversarial attacks such as evasion and data poisoning. The review emphasises the importance of detecting zero-day attacks, unknown malicious behaviours, and dynamically evolving threats, yet current defences fail to generalise across different attack scenarios. This limitation underscores the need for adaptive and explainable adversarial defence mechanisms that operate effectively in non-stationary network environments. Recent empirical studies further demonstrate the practical threats of adversarial attacks on DL-based NIDS. Qiu et al. [43] successfully bypassed Kitsune, a state-of-the-art NIDS, using model extraction and saliency maps, achieving a 94.31% attack success rate while modifying less than 0.005% of bytes in malicious packets. This finding reveals the high susceptibility of DL-based NIDS to adversarial evasion techniques, even under black-box settings where attackers lack full access to the target model. Similarly, authors in study [25] reviewed adversarial learning against DL-based NIDS, categorising attack types and assessing existing defences. Their findings indicate that current adversarial training strategies are often ineffective against novel adversarial variants, especially when deployed in large-scale SDN-IoT networks.

Furthermore, Zhang et al. [66] proposed TIKI-TAKA, a defence framework designed to enhance the robustness of DL-based NIDS against adversarial attacks. Their approach integrates model voting ensembling, adversarial training, and query detection, demonstrating improved resilience to adversarial perturbations. However, adversarial training remains computationally expensive and often fails to generalise across different types of attacks. The reliance on static adversarial training datasets limits the framework's effectiveness in real-world SDN-IoT networks, where attack patterns continuously evolve. Sánchez et al. [47] explored adversarial robustness in IoT device identification, employing LSTM-CNN architectures against contextual and evasion attacks. Although effective for fingerprinting, the study focuses narrowly on device-level patterns and overlooks system-wide SDN-IoT threats. Similarly, Bommana et al. [12] proposed a hybrid RCNN-RBM model with adaptive filtering and feature optimisation for IoT security. While it demonstrates promising adversarial robustness, it is not validated on widely accepted benchmarks and does not address SDN controller-level vulnerabilities. Ennaji et al. [20] presented a comprehensive survey of adversarial challenges in ML-based NIDS, offering a valuable taxonomy of threats and defences. However, their discussion remains general and lacks a specific focus on DL-based AAD systems and the SDN-IoT domain.

These challenges underscore the need for more adaptive, efficient, low-overhead defence mechanisms. Despite growing research in this domain, several key challenges remain unaddressed. First, DL models in SDN-IoT networks remain highly vulnerable to adversarial attacks due to the lack of robust feature representations and interpretability mechanisms. Second, while adversarial training is explored as a defence, it suffers from significant computational overhead and poor adaptability to evolving attack strategies. Third, a lack of standardised adversarial benchmarks and evaluation metrics for SDN-IoT environments makes it difficult to assess and compare the effectiveness of different defence techniques. Additionally, existing defence strategies are often designed for computer vision tasks and fail to account for the unique characteristics of network traffic data, such as time-series dependencies and protocol-specific constraints.

Table 1: Summary of related work on adversarial attacks and defences in DL-based security systems for SDN-IoT and IoT environments

| Study | Year | Focus | Key Contributions | Gaps Identified | Key Areas Considered | | | |
|---|---|---|---|---|---|---|---|---|
| | | | | | Adv-Attacks in DL | Threat Models | AAD | IoT/ SDN/ SDN-IoT |
| Qiu et al. [43] | 2020 | Adversarial attacks on DL-based NIDS in IoT | Demonstrated compromise of Kitsune NIDS using model extraction and saliency maps | Focused on NIDS in IoT, not broader SDN-IoT contexts | ✓ | ✓ | ✗ | ✓ |
| Aversano et al. [8] | 2021 | DL methods for IoT and SDN-IoT security | Review of DL methods; highlight vulnerabilities to adversarial sampling and data poisoning | Lack of countermeasures; need for detection of zero-day attacks and dynamic threats | ✗ | ✗ | ✗ | ✓ |
| Chakraborty et al. [15] | 2021 | Adversarial attacks and defences in DL | Survey of adversarial attacks; discussion on the need for robust DL models | General focus on attacks and defences for DL models, not specific to SDN-IoT domain | ✓ | ✓ | ✗ | ✗ |
| Alsoufi et al. [5] | 2021 | Anomaly detection in IoT using DL | Analysis of DL techniques for IoT intrusion detection | Does not specifically address adversarial attacks under SDN-IoT | ✗ | ✗ | ✓ | ✓ |
| Zhang et al. [66] | 2022 | DL-based NIDS and adversarial defenses | Proposed TIKI-TAKA framework and defense mechanisms like model voting ensembling | Focus on NIDS; need more strategies for AAD in SDN-IoT context | ✓ | ✗ | ✓ | ✓ |
| Ahmed et al. [1] | 2023 | ML-DL algorithms for SDN security | Comprehensive overview of algorithms for SDN security | Limited focus on adversarial attacks on DL models in SDN-IoT | ✗ | ✗ | ✗ | ✓ |
| Bai et al. [9] | 2023 | Adversarial robustness in continual learning | Identification of research gaps in adversarial settings; proposal of TABA for robustness | Specific focus on continual learning, limited to CIFAR-10 and CIFAR-100 datasets | ✓ | ✓ | ✗ | ✗ |
| Taheri et al. [49] | 2023 | DL for SDN and SDN-IoT security | Survey of DL applications and challenges in large-scale networks | Limited focus on adversarial attacks | ✗ | ✗ | ✗ | ✓ |
| He et al. [25] | 2023 | Adversarial ML for NIDS | Discussion of adversarial attacks (white-box and black-box), defences, and future research | General discussion, need for a more specific focus on AAD in SDN-IoT | ✓ | ✓ | ✗ | ✓ |
| Sánchez et al. [47] | 2024 | Adversarial ML for IoT fingerprinting | Device ID with LSTM-CNN, evaluated against context and evasion attacks | Focus on device-level ID, not full SDN-IoT systems | ✓ | ✓ | ✗ | ✓ |
| Ennaji et al. [20] | 2024 | Survey of adversarial challenges in ML-based NIDS | Comprehensive taxonomy of adversarial threats and defenses for NIDS | Focus on NIDS, not SDN-IoT-specific AAD | ✓ | ✓ | ✗ | ✓ |
| Bommana et al. [12] | 2025 | Adversarial attack defense in IoT using DL-AI | Hybrid RCNN-RBM model with adaptive filtering | Not evaluated on standard datasets; lacks SDN-IoT linkage | ✓ | ✓ | ✓ | ✗ |
| **Our Work (Proposed)** | 2025 | Comprehensive evaluation of adversarial attacks and defences in SDN-IoT | Benchmarking, threat model taxonomy, empirical evaluation, conceptual defense analysis | Real-world validation and comparative defense evaluation needed | ✓ | ✓ | ✓ | ✓ |

While prior surveys provide valuable insights, this study aims to bridge the gap by offering a comprehensive perspective on adversarial attacks against DL-based AAD systems in SDN-IoT networks. Our work focuses on the impact of adversarial attacks on DL-based AAD models, the effectiveness of adaptive adversarial defences, and the limitations of current solutions in real-world deployment scenarios. While existing surveys provide valuable insights, this SoK paper addresses the need for a more focused examination of adversarial attacks on DL-based AAD systems in SDN-IoT networks, offering a comprehensive perspective and filling a notable gap in the current literature.

## 4. Adversarial Attacks: Taxonomy and Analysis

This section analyses several adversarial attacks on DL-based AAD security systems deployed in SDN-IoT networks (See Figure 4). Specifically, this study addresses the following types of attacks.

- **White-Box Attacks** [56]: Fast Gradient Method (FGM) Attack, Carlini and Wagner (C&W) Attack, DeepFool Attack
- **Black-Box Attacks** [10]: Poisoning Attack, Membership Inference Attack
- **Grey-Box Attacks** [35]: Transferable Attack with FGM

For clarity, the underlying assumptions for each adversarial setting are stated as follows: in the **white-box** scenario, the adversary is assumed to have complete access to the model architecture, parameters, and training data; in the **grey-box** scenario, the adversary is considered to have partial knowledge, such as access to the data distribution but not the exact model weights; and in the **black-box** scenario, the adversary is assumed to have no visibility into the target model, instead relying on transferability from surrogate models under limited queries. These assumptions are consistent with standard practices in adversarial machine learning.

This paper examines the distinct characteristics and implications of these attacks at various levels, including data, model, and data-model level attacks. Our study focuses on those categorisations to better address the nature of adversarial attacks in SDN-IoT environments. In addition, white-box, black-box, and grey-box categorisations provide a comprehensive understanding of these attacks. Furthermore, a detailed discussion of the threat landscape for each attack type is presented using threat models. The section concludes with experimental findings, providing an in-depth understanding of the practical consequences of these attacks on security systems that leverage DL-based AAD models in the dynamic environment of SDN-IoT networks.

### 4.1. Attack Dimensions
#### 4.1.1. Data-Level Attacks

Data-level attacks focus on manipulating input data to mislead DL models. The **Fast Gradient Method (FGM) Attack** is a prominent attack technique in this category [23]. This attack perturbs input features based on the gradient of the loss function, exploiting the model's sensitivity to small changes in the input data. The adversarial perturbation ($\Delta$) is computed as follows:

$$\Delta = \epsilon \cdot \text{sign}(\nabla_x J(f(x), y_{\text{true}})) \tag{1}$$

where $\epsilon$ controls the magnitude of the perturbation, $f(x)$ is the model's prediction for input $x$, and $J$ is the loss function.

Another significant data-level attack is the **Poisoning Attack**, where adversaries insert malicious samples during the training phase [32]. This attack aims to compromise the model's learning process, resulting in biased or incorrect predictions when the model is exposed to poisoned data during inference. The modified loss function for poisoning can be expressed as:

$$J'(\theta) = J(\theta) + \lambda \sum_{i=1}^{N} \text{dist}(\mathbf{X}_{\text{orig}}, \mathbf{X}_{\text{poison},i}) \tag{2}$$

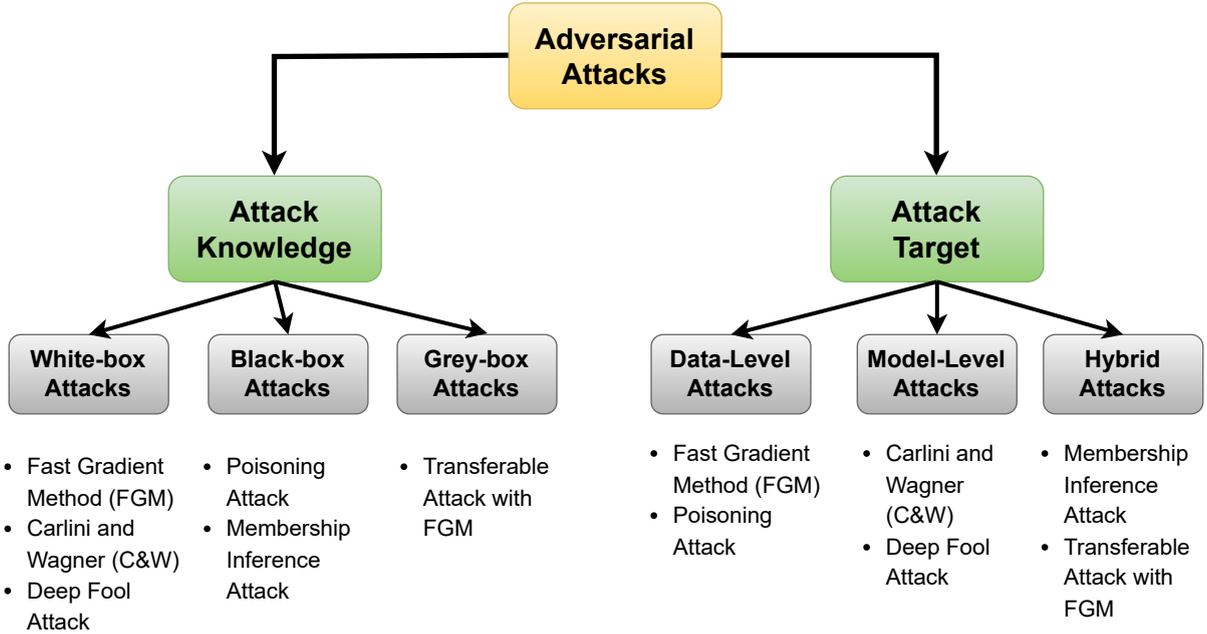

Figure 4: Taxonomy of adversarial attacks categorised by attack knowledge (White-box, Black-box, Grey-box) and attack target (Data-Level, Model-Level, Hybrid), with examples

where $J'(\theta)$ is the modified loss function and the original loss function $J(\theta)$ modifies by adding a regularization term. $\lambda$ controls the impact of the poisoning term, $N$ is the number of poisoning samples. The regularisation term penalises the distance between the original input data ($\mathbf{X}_{\text{orig}}$) and the injected poison data ($\mathbf{X}_{\text{poison},i}$).

### 4.1.2. Model-Level Attacks

Model-level attacks target the underlying DL model, seeking to exploit vulnerabilities in its architecture and decision-making process. The **DeepFool Attack** is an iterative technique designed to minimise the distance between the original and target predictions [38]. It achieves this by perturbing input features in an iterative manner, pushing them towards the decision boundary. The perturbation is computed as:

$$r = \arg\min_{r} \|\nabla_x f(x)^T r\|_2 \text{ subject to } \|\mathbf{r}\|_2 \leq \epsilon \tag{3}$$

The DeepFool attack aims to find the slightest perturbation $r$ such that the decision boundary of the model $f(x)$ is crossed. It minimises the norm of the perturbation subject to a constraint on its size ($\epsilon$).

On the other hand, the **Carlini and Wagner (C&W) Attack** is an optimisation-based approach that finds minimal perturbations to achieve a desired outcome [14]. C&W Attack formulates the optimisation problem:

$$\min_{r} \left( \|r\|_2^2 + c \cdot f(\mathbf{X} + \mathbf{r}) \right) \tag{4}$$

The C&W attack formulates the perturbation $r$ as a minimisation problem. It seeks to minimise the squared norm of the perturbation while ensuring that the perturbed input ($\mathbf{X} + \mathbf{r}$) is classified with a different target label ($c$). Despite being computationally expensive, the DeepFool and C&W attacks are known for their effectiveness against various defence mechanisms.

### 4.1.3. Data and Model Level Attacks

Hybrid attacks combine strategies from both data and model-level dimensions. **Transferable Attacks with FGM** involves creating adversarial examples that generalise across different models. It makes them particularly powerful, as the generated adversarial examples can fool diverse models, posing a significant challenge for defenses [39]. FGM transferability can be expressed as:

$$x' = x + \epsilon \cdot \text{sign}(\nabla_x J(f(x), y_{\text{true}}))  \quad (5)$$

Table 2: Threat model landscape for proposed attacks

| Threat Model Characteristics | Type | Attacker View | | | | | |
|---|---|---|---|---|---|---|---|
| | | Data Level | | Model Level | | Data and Model Level | |
| | | FGM Attack | Poisoning Attack | DeepFool Attack | C&W Attack | Transferable Attack with FGM | Membership Inference Attack |
| Attacker's Knowledge | Training data | ✓ | ✓ | ✗ | ✗ | ✓ | ✓ |
| | Feature set | ✓ | ✓ | ✓ | ✓ | ✓ | ✓ |
| | Feature extractor | ✓ | ✓ | ✓ | ✓ | ✓ | ✓ |
| | Feature transformers | ✓ | ✓ | ✓ | ✓ | ✓ | ✓ |
| | Parameters and hyper-parameters | ✗ | ✗ | ✓ | ✓ | ✓ | ✓ |
| | Objective function | ✗ | ✗ | ✓ | ✓ | ✓ | ✓ |
| | Inference API | ✗ | ✗ | ✓ | ✓ | ✓ | ✓ |
| | Model Behaviors | ✗ | ✗ | ✓ | ✓ | ✓ | ✓ |
| | Confidence Intervals | ✗ | ✗ | ✓ | ✓ | ✓ | ✓ |
| Attacker's Goal | Minimize perturbation | ✗ | ✗ | ✓ | ✓ | ✗ | ✗ |
| Attacker's Capability | | Depends on each specific attack | | | | | |
| Attacker's Strategy | Satisfy domain constraints | ✓ | ✗ | ✗ | ✗ | ✓ | ✓ |

It generates an adversarial example $x'$ by perturbing the input $x$ in the direction of the gradient of the loss function. *epsilon* scales the perturbation to control the magnitude.

The **Membership Inference Attack** also focuses on determining whether a specific data point was part of the training set [48]. This attack raises privacy concerns, as successful inference could reveal information about the training data, compromising the confidentiality of the model. A binary classifier $g$ is trained and the attack can be represented as:

$$g(x) = \sigma(f(x)) \quad (6)$$

This attack estimates the membership probability by applying a sigmoid function $\sigma$ to the model's output $f(x)$. The sigmoid function transforms the raw output into a probability score, indicating the likelihood that the input $x$ belongs to the training set.

4.2. Threat Models for Attacks

Adversarial attacks on DL-based AAD systems deployed in SDN-IoT networks pose distinct threats to the integrity of input data and the robustness of the underlying DL models. Understanding these threats is crucial for developing effective defenses [58]. Table 2 shows a combined threat model for the proposed attacks.

4.2.1. Threat Model for Data-Level Attacks

Adversaries launching data-level attacks possess knowledge of the DL model's architecture and the training dataset. Their primary goal is to manipulate the training data to introduce adversarial perturbations or inject poisoned samples. Capable of subtly altering a subset of the training data, their strategy involves introducing biases that mislead the DL model during the learning process. By exploiting vulnerabilities in the training data, attackers aim to influence the model's decision boundaries, which can potentially lead to misclassifications or biased predictions.

*4.2.2. Threat Model for Model-Level Attacks*

Attackers undertaking model-level attacks demonstrate a deep understanding of the DL model's architecture, parameters, and possibly its training process. They aim to compromise the model's decision boundaries by manipulating its parameters or injecting adversarial examples during testing. Proficient in crafting adversarial samples, these attackers employ optimisation-based techniques or generate misleading inputs to deceive the DL model. The strategy involves exploiting structural weaknesses, which may cause the model to make incorrect predictions due to adversarial inputs.

*4.2.3. Threat Model for Data and Model Level Attacks*

Data and model-level attacks need comprehensive knowledge of the training data and the DL model's architecture. An attacker aims to concurrently manipulate input data and exploit vulnerabilities in the DL model, creating a combined impact. Capable of executing strategies from both data-level and model-level attacks, these adversaries integrate techniques to manipulate training data and craft adversarial examples. This multifaceted approach yields a more robust and precise attack, thereby challenging the robustness of the DL model against adversarial scenarios.

*4.2.4. Threat Mapping to SDN-IoT Layers*

Adversarial threats affect different layers of the SDN-IoT architecture in distinct ways. **Data-level attacks**, such as FGM and poisoning, primarily impact the *data plane*, where IoT devices and edge nodes generate traffic. These attacks manipulate input features before classification, allowing malicious traffic to evade detection or corrupt the learning process [18]. **Model-level attacks**, including C&W and Deep-Fool, compromise the *application and control planes*, where DL models typically reside in SDN controllers or edge-based decision engines. Such attacks exploit model vulnerabilities by introducing carefully crafted perturbations that mislead classification outcomes. **Hybrid attacks**, such as transferable adversarial examples and membership inference, span multiple layers of the architecture. Transferable attacks exploit *model generalisation flaws*, affecting both edge and centralised nodes. In contrast, membership inference attacks may expose sensitive training data processed in the *control plane*, leading to potential privacy breaches [22]. Understanding this mapping helps identify which architectural components are most vulnerable and informs the deployment of targeted defensive strategies within SDN-IoT systems.

*4.3. Experiments and Findings*

*4.3.1. Experimental Setup*

In this SoK research, we assessed the robustness of the DL model in AAD systems against adversarial attacks within SDN-IoT environments. Our experiments employed three widely recognized and diverse datasets—CICIDS2017 [1], InSDN [2], and CICIoT2023 [3] (refer to Table 3). **These datasets were selected due to the current lack of publicly available, standardised datasets that fully capture SDN-IoT networks. Each dataset contributes unique characteristics reflective of the SDN-IoT domain:** *CICIDS2017* contains extensive traffic types and serves as a baseline for attack diversity in conventional networks; *InSDN* is tailored to SDN environments, capturing control-plane and data-plane behaviours; and *CICIoT2023* offers realistic IoT traffic scenarios and lightweight device interactions, closely simulating IoT endpoints. Collectively, they represent key operational layers and threat surfaces in SDN-IoT networks. This combination enables a comprehensive analysis of DL-based AAD robustness across multiple threat models and network configurations. Distributed Denial of Service (DDoS) and malware attacks were prioritised, given their prevalence and severity in SDN-IoT networks.

---

[1]https://www.unb.ca/cic/datasets/ids-2017.html
[2]https://aseados.ucd.ie/datasets/SDN/
[3]https://www.unb.ca/cic/datasets/iotdataset-2023.html

Table 3: Dataset descriptions

| Dataset | Total samples | Samples used | Features used | Labels used |
|---------|--------------|--------------|---------------|-------------|
| CICIDS2017 | 2,829,385 | 250,000 | 78 | DDoS, DoS, Bot, BENIGN |
| InSDN | 343,889 | 342,275 | 77 | DDoS, DoS, BOTNET, Normal |
| CICIoT2023 | 234,745 | 234,745 | 39 | DDoS, DoS, Mirai, Benign |

To address the detection of adversarial threats, we trained a modified Convolutional Neural Network (CNN) model tailored for anomaly detection tasks within SDN-IoT contexts (see Table 4). The architecture is designed to strike a balance between detection accuracy and computational efficiency. Before training, all datasets underwent standardised preprocessing steps, including normalisation of feature values to a [0, 1] range, one-hot encoding of categorical class labels, and removal of constant or irrelevant features to improve model generalisation and ensure input consistency. Given the computational cost associated with generating adversarial attacks, we adopted a single representative CNN model to maintain experimental feasibility while offering meaningful insights into adversarial vulnerability and defence. This design choice aligns with the goal of this SoK paper, which provides a reproducible, practical benchmark for evaluating DL robustness in diverse SDN-IoT threat environments.

Table 4: CNN Model Configuration Summary

| Component | Configuration |
|-----------|---------------|
| Conv Layers | 2 × Conv1D (64 → 128 filters, kernel size = 6, ReLU) |
| Pooling | MaxPooling1D (pool size = 2) after each Conv layer |
| Dense Layers | 2 × Dense (128 → 64 units, ReLU) + Dropout (rate = 0.1) |
| Output Layer | Dense ($n\_classes$ units, Softmax) |
| Training | Adam (learning rate = 0.0001), Categorical Cross-Entropy, 30 epochs, batch size = 128 |

We followed a standardised training procedure to ensure consistency and reliability. The dataset was partitioned in a 70:30 ratio for training and testing, with the testing set further divided equally for validation and final testing. This study employed precision, accuracy, F1-score, attack cost and recall metrics to assess the performance of the DL model and its vulnerability to adversarial attacks. All adversarial attacks discussed in previous sections were implemented and executed using the Adversarial Robustness Toolbox [4], following the default parameter configurations specified in its documentation to ensure consistency and reproducibility with prior adversarial machine learning studies. The experiments were conducted on a high-performance server equipped with dual Intel Xeon Gold 6140 CPUs (totalling 72 cores) and 256 GB of system memory, running a Linux-based operating system.

*4.3.2. Results and Discussion*

Based on the above experiments, we analysed the results (shown in Table 6) and their implications on the robustness of deep learning-based AAD systems in SDN-IoT networks. The findings provide valuable insights into the effectiveness of adversarial attacks, computational costs, and possible risks associated with deploying DL-based AAD models in SDN-IoT environments.

- **Fast Gradient Method (FGM) Attack:** The FGM attack resulted in a notable accuracy drop across all three datasets, demonstrating its effectiveness in fooling DL models with minimal perturbations. The accuracy dropped from 98.23% to 97.23% in CICIDS2017, 99.52% to 64.10% in InSDN, and 99.64% to 83.88% in CICIoT2023. The observed accuracy drop on the CICIDS2017 dataset for FGM attack results aligns with findings by Liu et al. [34]. In addition, results show that InSDN was the most vulnerable dataset, experiencing a substantial degradation in model accuracy. The attack was computationally efficient [42], with execution times ranging from 11.44s (CICIoT2023) to 19.13s (InSDN). The rapid nature of this attack makes it a serious threat in real-time SDN-IoT deployments, where attackers can quickly generate adversarial examples and evade detection. In a real-world SDN-IoT deployment, the efficiency of FGM poses a significant threat, as an attacker can quickly generate adversarial samples with minimal computational resources. It enables rapid evasion of security measures, particularly in IoT environments where lightweight models are frequently employed. Given its low computational cost, FGM attacks can be executed frequently, increasing the risk of real-time attack deployment in security-sensitive environments.

---

[4]https://github.com/Trusted-AI/adversarial-robustness-toolbox

Table 5: Server Configuration Details

| Component | Specification |
|---|---|
| CPU(s) | Intel Xeon Gold 6140 (x2) |
| Cores | 72 |
| Memory | 256GB |
| Operating System | Linux |

Table 6: Experiment results for the Fast Gradient Method (FGM) Attack, Poisoning Attack, DeepFool Attack, Carlini and Wagner (C&W) Attack, Transferable Attacks with FGM, and Membership Inference Attack on CICDS2017, InSDN and CICIoT2023 datasets

| Dataset | Metric | CNN before Attack | Data Level Attacks | | Model Level Attacks | | Data & Model Level Attacks | |
|---|---|---|---|---|---|---|---|---|
| | | | FGM | Poisoning | C&W | DeepFool | Transferable with FGM | Membership Inference |
| CICIDS2017 | Accuracy (%) | 98.23 | 97.23 | 96.36 | 93.45 | 94.11 | 97.37 | 93.70 |
| | Recall (%) | 98.23 | 97.23 | 96.36 | 93.45 | 94.11 | 97.37 | 93.70 |
| | Precision (%) | 98.25 | 97.28 | 96.37 | 93.57 | 95.66 | 97.46 | 99.99 |
| | F1 Measure (%) | 98.21 | 97.15 | 96.28 | 92.09 | 94.66 | 97.31 | 96.75 |
| | Time (seconds) | - | 15.35 | 0.47 | 8115.6 | 3489.6 | 18.85 | 456.12 |
| InSDN | Accuracy (%) | 99.52 | 64.10 | 94.17 | 78.53 | 92.04 | 72.19 | 51.12 |
| | Recall (%) | 99.52 | 64.10 | 94.17 | 78.53 | 92.04 | 72.19 | 51.12 |
| | Precision (%) | 99.57 | 85.17 | 94.42 | 74.18 | 93.43 | 87.30 | 99.99 |
| | F1 Measure (%) | 99.50 | 61.33 | 94.20 | 76.29 | 92.37 | 71.76 | 67.66 |
| | Time (seconds) | - | 19.13 | 0.206 | 7902.1 | 6324.3 | 25.87 | 1753.2 |
| CICIoT2023 | Accuracy (%) | 99.64 | 83.88 | 97.04 | 79.60 | 94.79 | 83.79 | 73.59 |
| | Recall (%) | 99.64 | 83.88 | 97.04 | 79.60 | 94.79 | 83.79 | 99.99 |
| | Precision (%) | 99.68 | 85.30 | 97.06 | 88.10 | 95.06 | 85.62 | 73.59 |
| | F1 Measure (%) | 99.66 | 78.28 | 97.05 | 81.58 | 94.86 | 78.20 | 84.79 |
| | Time (seconds) | - | 11.44 | 0.41 | 6801.6 | 325.8 | 15.33 | 366.24 |

- **Poisoning Attack:** The vulnerability of ML models to data poisoning attacks contains numerous prior research works and findings [2]. Experiment results demonstrate that poisoning attacks significantly degrade model accuracy across all datasets, highlighting the severe risk of data contamination during model training. The accuracy drop was 1.87% in CICIDS2017, 5.35% in InSDN, and 2.60% in CICIoT2023. The considerable reduction in accuracy and other metrics on the CICIDS2017, InSDN, and CICIoT2023 datasets aligns with the work of Dunn et al. [19]. The authors highlight the effectiveness of poisoning attacks in compromising model integrity using ToN-IoT and UNSW NB-15 datasets. While the reduction appears moderate compared to other attacks, the long-term impact of poisoning attacks is more alarming, as they compromise model learning, leading to persistent vulnerabilities. Unlike other attacks targeting inference-time performance, poisoning attacks corrupt the training phase, making them stealthy and difficult to detect. The execution time was exceptionally low (0.41s – 0.47s), meaning attackers can manipulate datasets with minimal computational resources. In practice, poisoning attacks are more dangerous in AAD systems that rely on continuous learning, where the model periodically updates itself using new data from network traffic. If an attacker can inject poisoned samples into the training pipeline, the model's accuracy will gradually deteriorate, resulting in the misclassification of malicious traffic as benign.

- **DeepFool Attack:** DeepFool proved to be a highly effective evasion attack, with accuracy dropping by 4.12% in CICIDS2017, 7.48% in InSDN, and 4.85% in CICIoT2023. The attack works by iteratively modifying input samples until they cross the decision boundary, making it a potent technique for fooling classifiers. However, the primary limitation of DeepFool is its high computational cost, which requires between 325.8 seconds (CICIoT2023) and 6,324.3 seconds (InSDN) to execute. Due to its high time cost, DeepFool is less practical for large-scale or real-time adversarial attacks [4]. However, it remains a powerful technique in targeted attacks, where adversaries possess sufficient resources and access to model parameters. Given enough time and access to offline model versions, an attacker can craft highly effective adversarial samples and later deploy them to compromise an AAD system in SDN-IoT networks. When analysed with existing works, these findings provide valuable insights into the characteristics of the DeepFool attack and its implications for deep learning model robustness [41].

- **Carlini and Wagner (C&W) Attack:** The C&W attack's significant reduction in model accuracy aligns with prior research demonstrating the effectiveness of optimisation-based attacks. These findings offer valuable insights into the impact of C&W attacks on model security. Notably, our results are consistent with the original work by Carlini and Wagner [14], showcasing the persistence of such attacks across diverse datasets. Among the evaluated attacks, C&W caused the most significant accuracy degradation, dropping from 98.23% to 93.45% in CICIDS2017, 99.52% to 78.53% in InSDN, and 99.64% to 79.60% in CICIoT2023. These results highlight the potency of optimisation-based adversarial attacks, which systematically craft perturbations that mislead the model while remaining imperceptible. The significant accuracy drop across datasets confirms that C&W poses a severe threat to model reliability, particularly in high-security environments where precision is crucial. However, despite its effectiveness, the C&W attack demands extensive computational resources, with execution times ranging from 6801.6s (CICIoT2023) to 8115.6s (CICIDS2017). This high computational cost makes it impractical for real-time adversarial attack scenarios. Nonetheless, C&W remains a formidable tool in targeted adversarial campaigns where attackers have sufficient time to generate optimised adversarial examples [41]. Its ability to bypass robust security measures underscores the critical challenge of deploying DL-based AAD systems.

- **Transferable Attacks with FGM:** The transferable nature of adversarial examples was confirmed through this experiment, showing that adversarial samples crafted for one model were also effective against other models [42]. The attack resulted in accuracy drops of 0.86% in CICIDS2017, 27.33% in InSDN, and 15.85% in CICIoT2023. Notably, the InSDN dataset was highly vulnerable, indicating that specific dataset characteristics influence adversarial transferability. The efficient computational cost (15.33s – 25.87s) makes this attack a serious real-world concern since attackers do not need to tailor adversarial samples for each specific model. Instead, they can generate adversarial inputs on one system and deploy them against others, making cross-model security a key challenge for DL-based AAD systems.

- **Membership Inference Attack:** The membership inference attack caused the most severe accuracy drop, particularly in the InSDN dataset, where accuracy dropped by 48.40%. In CICIDS2017 and CICIoT2023, accuracy reductions were 4.53% and 26.05%, respectively. This attack poses a critical privacy risk, as adversaries can determine whether a particular data point was part of the model's training set [63]. Practically, it is like someone attempting to identify whether a specific network communication instance was part of the model's training set. The computational cost of the attack varied significantly, ranging from 366.24 seconds (CICIoT2023) to 1,753.2 seconds (InSDN). It suggests that datasets with lower variability may be more susceptible to membership inference attacks. The ability of an attacker to extract training data information poses a serious concern for privacy in DL-based security models, especially when handling sensitive network traffic data in SDN-IoT environments [61].

*4.3.3. Comparison and Key Insights*

The following Table 7 summarises the impact of each attack across datasets, highlighting the most vulnerable dataset, the most significant accuracy drop, and execution time:

Table 7: Comparison of Adversarial Attacks(Accuracy Drop, Affected Dataset, and Execution Time)

| Attack | Largest Accuracy Drop (%) | Most Affected Dataset | Execution Time (s) |
| --- | --- | --- | --- |
| FGM | ↓ 35.42% | InSDN | 11.44s – 19.13s |
| Poisoning | ↓ 5.35% | InSDN | 0.41s – 0.47s |
| DeepFool | ↓ 7.48% | InSDN | 325.8s – 6324.3s |
| C&W | ↓ 20.99% | InSDN | 6801.6s – 8115.6s |
| Transferable FGM | ↓ 27.33% | InSDN | 15.33s – 25.87s |
| Membership Inference | ↓ 48.40% | InSDN | 366.24s – 1753.2s |

These results highlight that the InSDN dataset was the most vulnerable across all attack types, suggesting that specific dataset characteristics influence susceptibility to adversarial attacks. While C&W and DeepFool attacks showed the highest impact, they were computationally expensive, making them less practical for real-time attacks. On the other hand, FGM, poisoning, and transferable attacks were computationally efficient, posing a more immediate threat to deployed AAD models. These findings emphasise the need for robust adversarial defences to safeguard DL-based AAD systems in SDN-IoT networks, ensuring security and model integrity against evolving adversarial threats.

Among the tested attack vectors, the FGM attack demonstrated the most dramatic degradation in model accuracy on the InSDN dataset (↓35.42%), compared to CICIDS2017 (↓1.00%) and CICIoT2023 (↓15.76%). This noticeable drop can be attributed to several unique characteristics of the InSDN dataset. First, InSDN includes detailed SDN-specific flow features related to the control and data planes, introducing high-dimensional input patterns susceptible to gradient-based perturbations. Second, the InSDN dataset exhibits relatively lower feature redundancy and more substantial class interdependence, which reduces the decision boundary margin for the classifier and amplifies its vulnerability to small input perturbations. Moreover, the richer set of protocol-specific attributes may have induced overfitting during training, making the model more fragile under adversarial influence. These factors explain why the same FGM perturbation strategy leads to significantly worse performance on InSDN than on the other datasets, highlighting the critical need for tailored defence strategies when dealing with SDN-focused traffic flows.

## 5. Defence, Mitigation, and Drawbacks of Adversarial Attacks

*5.1. Defence and Mitigation Strategies*

*5.1.1. Data-Level*

- **Data Sanitisation** is a data-level defence that involves preprocessing training data to eliminate potential adversarial samples or outliers, thereby enhancing model robustness. However, balancing effective sanitisation with preserving dataset diversity presents challenges, as overly aggressive cleansing can lead to data loss [29].

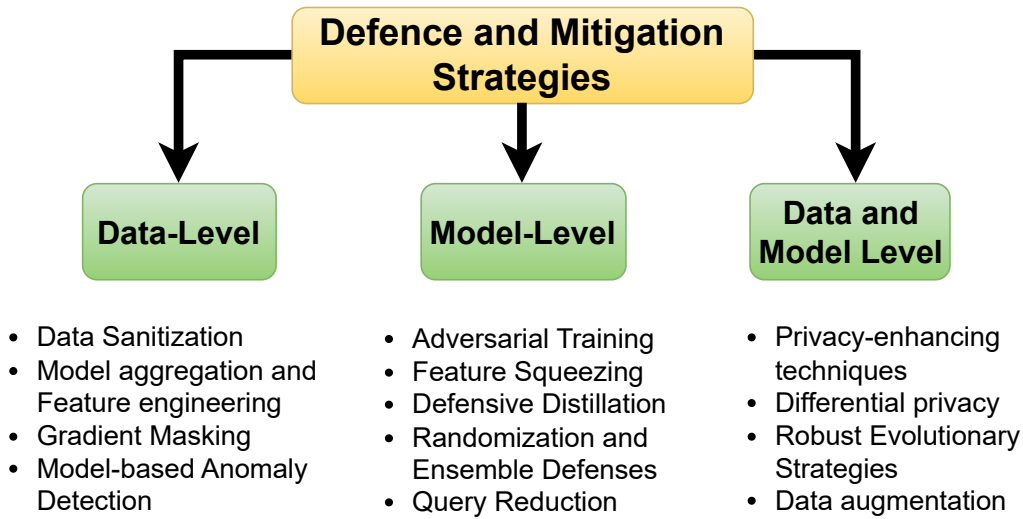

Figure 5: Defence and Mitigation Strategies against Adversarial Attacks in SDN-IoT Networks

- **Model aggregation and Feature engineering** form a collective defence strategy at the data level, combining predictions from multiple models and transforming input features, strengthening models against adversarial attacks. This approach mitigates the impact of attacks targeting specific vulnerabilities in individual models [53]. Salman et al. explored the impact of different aggregation strategies on the ensemble's stability and accuracy [46].

- **Gradient Masking** blurs gradient information during training, restricting adversaries from exploiting model sensitivity. However, it can negatively impact model performance, reducing accuracy and increasing training time [7, 54].

- **Model-based Anomaly Detection** involves specialised models trained to identify anomalous or adversarial instances; this data-level defence provides additional protection. However, its effectiveness relies on accurate training to capture the actual distribution of normal data [40, 3].

*5.1.2. Model-Level*
- **Adversarial Training** is an early and robust defence mechanism that enhances model robustness against adversarial attacks. Despite its proven effectiveness, adversarial training may increase training time and computational complexity and is only sometimes effective against all attack types [33, 36].

- **Feature Squeezing**, a model-level defence reducing the input space susceptible to adversarial perturbations. While effective against some attacks, it can reduce the accuracy of clean data [60, 64].

- **Defensive Distillation** involves training a model to emulate a pre-existing one and has shown effectiveness against specific adversarial attacks. However, it may reduce the accuracy of clean data and is only sometimes effective against all attacks [26, 51, 57].

- **Randomisation and Ensemble Defences** introduce variability at the input and model levels. Despite their effectiveness, they may pose challenges such as increased training time and computational complexity [55, 50, 51].

- **Query Reduction** was introduced to strategically limit an adversary's access to the model's predictions by imposing constraints on the number of queries. This defence mechanism restricts attacks relying on repeated model queries, introducing a delicate trade-off between robustness and model performance [13, 6].

*5.1.3. Data and Model Level*
- **Privacy-enhancing techniques** are crucial for securing machine learning systems against privacy breaches and adversarial threats. These techniques protect sensitive information, ensuring responsible and secure deployment of machine learning models. Privacy preservation is crucial when models handle sensitive datasets [59].

- **Differential privacy** is an ideal technique to ensure robust confidentiality by introducing controlled noise during training. It prevents adversaries from extracting sensitive information about individual data points, allowing models to be trained on sensitive datasets without compromising confidentiality [11, 31].

- **Robust Evolutionary Strategies** propose using evolutionary algorithms to optimise model parameters in pursuit of adversarial resilience. This evolutionary approach iteratively refines models to enhance their resistance against adversarial attacks [62, 52, 36].

- **Data augmentation** is an effective defence strategy at the intersection of data and model levels, synthetically expanding training datasets. Data augmentation improves model performance, contributes to privacy preservation, and enhances model generalisation and resilience to adversarial attacks [21, 65].

As this work is positioned as a Systematisation of Knowledge (SoK), our primary objective is to analyse and synthesise adversarial threats and defence strategies conceptually, focusing on their relevance to DL-based AAD in SDN-IoT networks. While many of the discussed techniques—such as adversarial training [36], feature squeezing [60], and differential privacy [31]—have demonstrated empirical success in prior research (See Table 8). Conducting experimental benchmarking across multiple datasets and attack vectors was beyond the intended scope of this study. Instead, we concentrate on assessing their theoretical strengths, limitations, and deployment feasibility within SDN-IoT environments. This systematised evaluation provides a foundation for future empirical studies that can build on these findings by validating selected defences under specific operational and adversarial conditions.

*5.2. Drawbacks of Adversarial Attacks*

While posing significant challenges, adversarial attacks come with inherent drawbacks that shape their limitations. One important constraint is the restraint to the L2 norm, which narrows the scope of attacks and may overlook vulnerabilities in alternative norms. Moreover, the requirement for adversaries to retain knowledge of the target model's architecture and parameters significantly limits the feasibility of attacks. This necessity for model information increases the complexity of attacks and poses a barrier, making specific attacks difficult without a deep understanding of the model's internals.

In addition to these model-specific limitations, adversarial attacks face computational costs and scalability challenges. Their resource-intensive nature renders them computationally expensive, particularly in scenarios with large-scale or real-time applications. This limitation raises practical concerns, as the scalability of adversarial attacks becomes a critical factor in their real-world applicability. The lack of diversity in attack strategies, sensitivity to hyperparameters, and limited transferability across different models further underscores the exact landscape of adversarial attacks, emphasising the need for a comprehensive understanding of their drawbacks for effective defence strategies.

*5.3. Feasibility in Edge and Resource-Constrained IoT Environments*

Real-world deployment of DL-based AAD systems requires careful consideration of the limitations inherent to edge computing environments and resource-constrained IoT devices. Even relatively compact

Table 8: Comparative Summary of Adversarial Defences for DL-based AAD in SDN-IoT

| Defense Strategy | Attack Coverage | Computational Cost | SDN-IoT Suitability | Empirical Support |
|---|---|---|---|---|
| Adversarial Training | High | High | Moderate (edge-unfriendly) | [36, 33] |
| Feature Squeezing | Moderate | Low | High (lightweight) | [60, 64] |
| Differential Privacy | High | Medium | Moderate | [31, 11] |
| Defensive Distillation | Low–Moderate | Medium | Moderate | [26, 51, 57] |
| Randomisation and Ensemble Defenses | Moderate | High | Low (resource-intensive) | [50, 55] |
| Query Reduction† | Black-box only | Low | High | [13, 6] |
| Data Sanitisation | Moderate | Low | High | [29] |
| Model Aggregation and Feature Engineering | Moderate | Medium | Moderate | [53, 46] |
| Gradient Masking | Low–Moderate | Medium | Moderate | [7, 54] |
| Model-based Anomaly Detection | Moderate | Medium | High | [40, 3] |
| Privacy-Enhancing Techniques | High | Medium | Moderate | [59] |
| Robust Evolutionary Strategies | High | High | Low–Moderate | [62, 52] |
| Data Augmentation | Moderate | Low | High | [21, 65] |

†Query Reduction defends only against black-box attacks that rely on repeated access to model outputs.

CNN architectures may impose significant overhead on devices with limited memory, processing capacity, and energy budgets. Model compression techniques such as pruning, quantisation, and lightweight CNN variants have been shown to improve feasibility without substantial loss of detection accuracy. Furthermore, edge–cloud collaborative frameworks provide a practical pathway, allowing lightweight feature extraction and initial AAD to occur at the edge. At the same time, more complex inference and adversarial analysis can be offloaded to nearby gateways or controllers. However, computationally intensive defence strategies, such as adversarial training, may remain challenging to realise directly on constrained devices. These trade-offs highlight the need for resource-aware adversarial defence mechanisms designed explicitly for SDN-IoT environments. This discussion also connects with Section 6.2, where integration with edge computing is identified as a future research direction.

## 6. Limitations and Future Research Directions

*6.1. Current Limitations*

Although this study provides a structured analysis of adversarial threats and defences for DL-based AAD in SDN-IoT networks, certain limitations should be acknowledged:

- **Single Model Focus:** The evaluation was conducted using a CNN-based architecture only, which may not fully capture the behaviour of other DL models such as RNNs, LSTMs, or hybrids.

- **Computational Assumptions:** Experiments were carried out on high-performance servers, and therefore may not reflect the computational constraints of real-time or edge-based IoT deployments.

- **Offline Dataset Limitation:** The evaluation relied on benchmark datasets (CICIDS2017, InSDN, CICIoT2023) in offline settings, and thus, real-time deployment aspects remain unexplored.

Recognising these limitations helps clarify the scope of this work and directly motivates the future research directions outlined in Section 6.2.

*6.2. Future Research Directions*

As the landscape of DL-based AAD in SDN-IoT networks continues to evolve, several approaches for future research can be listed as follows:

- **Real-time Adaptive Defence Mechanisms for AAD:** Investigate and develop real-time adaptive defence mechanisms for AAD, dynamically adjusting to evolving adversarial threats in SDN-IoT networks while ensuring swift responses.

- **Explainable AI in AAD for Adversarial Contexts:** Explore advancements in explainable AI focused on AAD systems, enhancing the interpretability of DL models during adversarial attacks, thereby aiding in understanding attack strategies.

- **Integration with Edge Computing:** Examine the implications of integrating DL-based AAD systems with edge computing for SDN-IoT networks, focusing on optimising performance and reducing latency.

- **Human-in-the-Loop Collaborative Defence for AAD:** Integrate human-in-the-loop defences specifically designed for AAD systems, facilitating collaboration between automated AAD capabilities and human insights to strengthen defence against adversarial attacks.

- **Resource-Efficient AAD Architectures** Focus on creating resource-efficient AAD architectures suitable for SaaS deployment in resource-constrained IoT devices, ensuring practicality in real-world scenarios.

# 7. Conclusion

This study systematically analysed adversarial threats targeting DL-based AAD systems in SDN-IoT networks. Through an extensive evaluation of popular benchmark datasets (CICIDS2017, InSDN, and CICIoT2023), we demonstrated that adversarial attacks significantly compromise the performance of AAD models, with detection accuracy reductions of up to 48.4% in some cases. Membership Inference Attacks exhibited the highest success rate in compromising model integrity, while C&W and DeepFool attacks proved highly effective in evasion. Our findings underscore the necessity for robust defensive mechanisms to safeguard SDN-IoT networks against adversarial threats. From a defensive perspective, adversarial training can improve model resilience but at a high computational cost, making real-time deployment challenging. Other mitigation techniques, such as feature squeezing, model aggregation, and ensemble methods, provided partial protection but struggled against adaptive attacks. These results underscore the need for more dynamic and resource-efficient adversarial defence mechanisms tailored to SDN-IoT environments.

Despite these insights, our study also identifies open research challenges, including the trade-offs between model robustness and computational efficiency, the need for privacy-preserving adversarial defences, and the role of XAI in enhancing security transparency. Future work should focus on developing real-time adaptive security frameworks that dynamically adjust to evolving adversarial threats while maintaining computational feasibility for large-scale SDN-IoT deployments. This study provides a critical reference for advancing the security of DL-based AAD systems by systematically categorising adversarial threats, evaluating their impact, and assessing countermeasures. Our findings highlight the importance of continuous research and innovation in advancing the security and robustness of DL-based AAD security systems for deployment within SDN-IoT networks.

**Acknowledgment**

This work did not get any external funding.

# References

[1] Ahmed, M.R., Shatabda, S., Islam, A.M., Robin, M.T.I., et al., 2023. Intrusion detection system in software-defined networks using machine learning and deep learning techniques–a comprehensive survey. Authorea Preprints .

[2] Alatwi, H.A., Aldweesh, A., 2021. Adversarial black-box attacks against network intrusion detection systems: A survey, in: 2021 IEEE World AI IoT Congress (AIIoT), IEEE. pp. 0034–0040.

[3] Alghanmi, N., Alotaibi, R., Buhari, S.M., 2022. Machine learning approaches for anomaly detection in iot: an overview and future research directions. Wireless Personal Communications 122, 2309–2324.

[4] Alotaibi, A., Rassam, M.A., 2023. Enhancing the sustainability of deep-learning-based network intrusion detection classifiers against adversarial attacks. Sustainability 15, 9801.

[5] Alsoufi, M.A., Razak, S., Siraj, M.M., Nafea, I., Ghaleb, F.A., Saeed, F., Nasser, M., 2021. Anomaly-based intrusion detection systems in iot using deep learning: A systematic literature review. Applied sciences 11, 8383.

[6] Andriushchenko, M., Croce, F., Flammarion, N., Hein, M., 2020. Square attack: a query-efficient black-box adversarial attack via random search, in: European conference on computer vision, Springer. pp. 484–501.

[7] Athalye, A., Engstrom, L., Ilyas, A., Kwok, K., 2018. Synthesizing robust adversarial examples, in: International conference on machine learning, PMLR. pp. 284–293.

[8] Aversano, L., Bernardi, M.L., Cimitile, M., Pecori, R., 2021. A systematic review on deep learning approaches for iot security. Computer Science Review 40, 100389.

[9] Bai, T., Chen, C., Lyu, L., Zhao, J., Wen, B., 2023a. Towards adversarially robust continual learning, in: ICASSP 2023-2023 IEEE International Conference on Acoustics, Speech and Signal Processing (ICASSP), IEEE. pp. 1–5.

[10] Bai, Y., Wang, Y., Zeng, Y., Jiang, Y., Xia, S.T., 2023b. Query efficient black-box adversarial attack on deep neural networks. Pattern Recognition 133, 109037.

[11] Baraheem, S., Yao, Z., 2022. A survey on differential privacy with machine learning and future outlook. arXiv preprint arXiv:2211.10708 .

[12] Bommana, S.R., Veeramachaneni, S., Ershad, S., Srinivas, M., 2025. Addressing adversarial attacks in iot using deep learning ai models. IEEE Access .

[13] Carlini, N., Athalye, A., Papernot, N., Brendel, W., Rauber, J., Tsipras, D., Goodfellow, I., Madry, A., Kurakin, A., 2019. On evaluating adversarial robustness. arXiv preprint arXiv:1902.06705 .

[14] Carlini, N., Wagner, D., 2017. Towards evaluating the robustness of neural networks, in: 2017 ieee symposium on security and privacy (sp), Ieee. pp. 39–57.

[15] Chakraborty, A., Alam, M., Dey, V., Chattopadhyay, A., Mukhopadhyay, D., 2021. A survey on adversarial attacks and defences. CAAI Transactions on Intelligence Technology 6, 25–45.

[16] Cherian, M., Varma, S.L., 2023. Secure sdn–iot framework for ddos attack detection using deep learning and counter based approach. Journal of Network and Systems Management 31, 54.

[17] Cook, A.A., Mısırlı, G., Fan, Z., 2019. Anomaly detection for iot time-series data: A survey. IEEE Internet of Things Journal 7, 6481–6494.

[18] Das, T., Shukla, R.M., Sengupta, S., 2024. Poisoning the well: Adversarial poisoning on ml-based software-defined network intrusion detection systems. IEEE Transactions on Network Science and Engineering .

[19] Dunn, C., Moustafa, N., Turnbull, B., 2020. Robustness evaluations of sustainable machine learning models against data poisoning attacks in the internet of things. Sustainability 12, 6434.

[20] Ennaji, S., De Gaspari, F., Hitaj, D., Kbidi, A., Mancini, L.V., 2024. Adversarial challenges in network intrusion detection systems: Research insights and future prospects. arXiv preprint arXiv:2409.18736 .

[21] Fonseca, J., Bacao, F., 2022. Research trends and applications of data augmentation algorithms. arXiv preprint arXiv:2207.08817 .

[22] Gadallah, W.G., Ibrahim, H.M., Omar, N.M., 2024. A deep learning technique to detect distributed denial of service attacks in software-defined networks. Computers & Security 137, 103588.

[23] Goodfellow, I.J., Shlens, J., Szegedy, C., 2014. Explaining and harnessing adversarial examples. arXiv preprint arXiv:1412.6572 .

[24] Hariharan, A., Gupta, A., Pal, T., 2020. Camlpad: Cybersecurity autonomous machine learning platform for anomaly detection, in: Advances in Information and Communication: Proceedings of the 2020 Future of Information and Communication Conference (FICC), Volume 2, Springer. pp. 705–720.

[25] He, K., Kim, D.D., Asghar, M.R., 2023. Adversarial machine learning for network intrusion detection systems: a comprehensive survey. IEEE Communications Surveys & Tutorials .

[26] Hinton, G., Vinyals, O., Dean, J., 2015. Distilling the knowledge in a neural network. arXiv preprint arXiv:1503.02531 .

[27] Javeed, D., Gao, T., Khan, M.T., 2021. Sdn-enabled hybrid dl-driven framework for the detection of emerging cyber threats in iot. Electronics 10, 918.

[28] Khalid, M., Hameed, S., Qadir, A., Shah, S.A., Draheim, D., 2023. Towards sdn-based smart contract solution for iot access control. Computer Communications 198, 1–31.

[29] Koh, P.W., Steinhardt, J., Liang, P., 2022. Stronger data poisoning attacks break data sanitization defenses. Machine Learning , 1–47.

[30] Kuppa, A., Grzonkowski, S., Asghar, M.R., Le-Khac, N.A., 2019. Black box attacks on deep anomaly detectors, in: Proceedings of the 14th international conference on availability, reliability and security, pp. 1–10.

[31] Lecuyer, M., Atlidakis, V., Geambasu, R., Hsu, D., Jana, S., 2019. Certified robustness to adversarial examples with differential privacy, in: 2019 IEEE symposium on security and privacy (SP), IEEE. pp. 656–672.

[32] Li, D., Li, Q., Ye, Y., Xu, S., 2020. Sok: Arms race in adversarial malware detection. arXiv preprint arXiv:2005.11671 .

[33] Li, L., Xie, T., Li, B., 2023. Sok: Certified robustness for deep neural networks, in: 2023 IEEE symposium on security and privacy (SP), IEEE. pp. 1289–1310.
[34] Liu, J., Nogueira, M., Fernandes, J., Kantarci, B., 2021. Adversarial machine learning: A multilayer review of the state-of-the-art and challenges for wireless and mobile systems. IEEE Communications Surveys & Tutorials 24, 123–159.
[35] Liu, Z., Luo, Y., Wu, L., Li, S., Liu, Z., Li, S.Z., 2022. Are gradients on graph structure reliable in gray-box attacks?, in: Proceedings of the 31st ACM International Conference on Information & Knowledge Management, pp. 1360–1368.
[36] Madry, A., Makelov, A., Schmidt, L., Tsipras, D., Vladu, A., 2017. Towards deep learning models resistant to adversarial attacks. arXiv preprint arXiv:1706.06083 .
[37] Mishra, P., Puthal, D., Tiwary, M., Mohanty, S.P., 2019. Software defined iot systems: Properties, state of the art, and future research. IEEE Wireless Communications 26, 64–71.
[38] Moosavi-Dezfooli, S.M., Fawzi, A., Frossard, P., 2016. Deepfool: a simple and accurate method to fool deep neural networks, in: Proceedings of the IEEE conference on computer vision and pattern recognition, pp. 2574–2582.
[39] Muncsan, T., Kiss, A., 2021. Transferability of fast gradient sign method, in: Intelligent Systems and Applications: Proceedings of the 2020 Intelligent Systems Conference (IntelliSys) Volume 2, Springer. pp. 23–34.
[40] Nassif, A.B., Talib, M.A., Nasir, Q., Dakalbab, F.M., 2021. Machine learning for anomaly detection: A systematic review. Ieee Access 9, 78658–78700.
[41] Nguyen, X.H., Nguyen, X.D., Le, K.H., 2022. Preventing adversarial attacks against deep learning-based intrusion detection system, in: International Conference on Information Security Practice and Experience, Springer. pp. 382–396.
[42] Pawlicki, M., Choraś, M., Kozik, R., 2020. Defending network intrusion detection systems against adversarial evasion attacks. Future Generation Computer Systems 110, 148–154.
[43] Qiu, H., Dong, T., Zhang, T., Lu, J., Memmi, G., Qiu, M., 2020. Adversarial attacks against network intrusion detection in iot systems. IEEE Internet of Things Journal 8, 10327–10335.
[44] Ren, K., Zheng, T., Qin, Z., Liu, X., 2020. Adversarial attacks and defenses in deep learning. Engineering 6, 346–360.
[45] Said Elsayed, M., Le-Khac, N.A., Dev, S., Jurcut, A.D., 2020. Network anomaly detection using lstm based autoencoder, in: Proceedings of the 16th ACM Symposium on QoS and Security for Wireless and Mobile Networks, pp. 37–45.
[46] Salman, R., Alzaatreh, A., Sulieman, H., 2022. The stability of different aggregation techniques in ensemble feature selection. Journal of Big Data 9, 1–23.
[47] Sánchez, P.M.S., Celdrán, A.H., Bovet, G., Pérez, G.M., 2024. Adversarial attacks and defenses on ml-and hardware-based iot device fingerprinting and identification. Future Generation Computer Systems 152, 30–42.
[48] Shokri, R., Stronati, M., Song, C., Shmatikov, V., 2017. Membership inference attacks against machine learning models, in: 2017 IEEE symposium on security and privacy (SP), IEEE. pp. 3–18.
[49] Taheri, R., Ahmed, H., Arslan, E., 2023. Deep learning for the security of software-defined networks: a review. Cluster Computing 26, 3089–3112.
[50] Tramèr, F., Kurakin, A., Papernot, N., Goodfellow, I., Boneh, D., McDaniel, P., 2017a. Ensemble adversarial training: Attacks and defenses. arXiv preprint arXiv:1705.07204 .
[51] Tramèr, F., Papernot, N., Goodfellow, I., Boneh, D., McDaniel, P., 2017b. The space of transferable adversarial examples. arXiv preprint arXiv:1704.03453 .
[52] Vargas, D.V., Kotyan, S., IIIT-NR, S., 2019. Evolving robust neural architectures to defend from adversarial attacks. arXiv preprint arXiv:1906.11667 3.
[53] Verdonck, T., Baesens, B., Óskarsdóttir, M., vanden Broucke, S., 2021. Special issue on feature engineering editorial. Machine Learning , 1–12.
[54] Wang, J., Zhang, H., 2019. Bilateral adversarial training: Towards fast training of more robust models against adversarial attacks, in: Proceedings of the IEEE/CVF international conference on computer vision, pp. 6629–6638.
[55] Wang, X., Li, J., Kuang, X., Tan, Y.a., Li, J., 2019. The security of machine learning in an adversarial setting: A survey. Journal of Parallel and Distributed Computing 130, 12–23.
[56] Wang, Y., Liu, J., Chang, X., Rodríguez, R.J., Wang, J., 2022. Di-aa: An interpretable white-box attack for fooling deep neural networks. Information Sciences 610, 14–32.
[57] Wu, B., Pan, H., Shen, L., Gu, J., Zhao, S., Li, Z., Cai, D., He, X., Liu, W., 2021. Attacking adversarial attacks as a defense. arXiv preprint arXiv:2106.04938 .
[58] Xiong, W., Lagerström, R., 2019. Threat modeling–a systematic literature review. Computers & security 84, 53–69.
[59] Xu, R., Baracaldo, N., Joshi, J., 2021. Privacy-preserving machine learning: Methods, challenges and directions. arXiv preprint arXiv:2108.04417 .
[60] Xu, W., Evans, D., Qi, Y., 2017. Feature squeezing: Detecting adversarial examples in deep neural networks. arXiv preprint arXiv:1704.01155 .
[61] Yan, H., Li, X., Li, H., Li, J., Sun, W., Li, F., 2021. Monitoring-based differential privacy mechanism against query flooding-based model extraction attack. IEEE Transactions on Dependable and Secure Computing 19, 2680–2694.
[62] Yao, C., Bielik, P., Tsankov, P., Vechev, M., 2021. Automated discovery of adaptive attacks on adversarial defenses. Advances in Neural Information Processing Systems 34, 26858–26870.
[63] Yao, Y., 2022. Exploration of membership inference attack on convolutional neural networks and its defenses, in: 2022 International Conference on Image Processing, Computer Vision and Machine Learning (ICICML), IEEE. pp. 604–610.
[64] Zantedeschi, V., Nicolae, M.I., Rawat, A., 2017. Efficient defenses against adversarial attacks, in: Proceedings of the 10th ACM Workshop on Artificial Intelligence and Security, pp. 39–49.
[65] Zeng, Y., Qiu, H., Memmi, G., Qiu, M., 2020. A data augmentation-based defense method against adversarial attacks in neural networks, in: Algorithms and Architectures for Parallel Processing: 20th International Conference, ICA3PP 2020, New York City, NY, USA, October 2–4, 2020, Proceedings, Part II 20, Springer. pp. 274–289.

[66] Zhang, C., Costa-Perez, X., Patras, P., 2022. Adversarial attacks against deep learning-based network intrusion detection systems and defense mechanisms. IEEE/ACM Transactions on Networking 30, 1294–1311.